\documentstyle[12pt,aaspp4]{article} 
 
\slugcomment{ApJ, in press} 
 
\begin{document} 

\title{Energy Diagnoses of Nine Infrared Luminous Galaxies 
Based on 3--4 Micron Spectra}
\author{Masatoshi Imanishi \altaffilmark{1}}  
\affil{National Astronomical Observatory, Mitaka, Tokyo 181-8588, Japan}
\author{and \\
C. C. Dudley \altaffilmark{2}}  
\affil{Naval Research Laboratory, 
Remote Sensing Division, 
Code 7217, Building 2, Room 240B, 
4555 Overlook Ave SW, Washington DC 20375-5351, U.S.A.}
 
\altaffiltext{1}{Institute for Astronomy, 
University of Hawaii, 2680 Woodlawn Drive, Honolulu, Hawaii 96822, 
USA}

\altaffiltext{2}{NRC-NRL Research Associate}

\begin{abstract} 
 
The energy sources of nine infrared luminous galaxies (IRLGs)   
are diagnosed based on their ground-based 3--4 $\mu$m 
spectra.
Both the equivalent width of the 3.3 $\mu$m polycyclic aromatic 
hydrocarbon (PAH) emission feature and 
the 3.3 $\mu$m PAH to far-infrared luminosity ratio 
($L_{3.3}$/$L_{\rm FIR}$) are analyzed. 
Assuming nuclear compact starburst activity in these sources 
produces the 3.3 $\mu$m PAH emission as strongly as that in 
starburst galaxies with lower far-infrared luminosities,   
the followings results are found: 
For six IRLGs, both the observed equivalent widths and the 
$L_{3.3}$/$L_{\rm FIR}$ ratios are too small to explain the bulk of 
their far-infrared luminosities by compact starburst 
activity, indicating that active galactic nucleus (AGN) activity 
is a dominant energy source. 
For the other three IRLGs, while the 3.3 $\mu$m PAH equivalent widths 
are within the range of starburst galaxies, the $L_{3.3}$/$L_{\rm FIR}$ 
ratios after correction for screen dust extinction 
are a factor of $\sim$3 smaller.
The uncertainty in the dust extinction correction factor and 
in the scatter of the intrinsic $L_{3.3}$/$L_{\rm FIR}$ ratios for 
starburst galaxies do not allow a determination of the ultimate energy 
sources for these three IRLGs.

\end{abstract} 
 
\keywords{galaxies: active --- galaxies: nuclei --- galaxies: starburst 
--- infrared: galaxies --- galaxies: individual (IC 694, Arp 220, 
NGC 6240, Mrk 273, Mrk 231, IRAS 05189$-$2524, IRAS 08572+3915, 
IRAS 23060+0505, IRAS 20460+1925, NGC 253)} 
 
\section{Introduction} 

Infrared luminous galaxies (IRLGs), most of whose huge 
luminosities are radiated in the far-infrared (40--500 $\mu$m) 
as dust emission   
($L_{\rm FIR}$ ${\ ^{\displaystyle >}_{\displaystyle \sim}\ }$ 
10$^{45}$ ergs s$^{-1}$),
\footnote{
$H_{0}$ $=$ 75 km s$^{-1}$ Mpc$^{-1}$ and $q_{0}$ $=$ 0.5 
are used throughout this paper. 
} 
are a significant population at the bright end of galaxy luminosity 
function (Soifer et al. 1987).
Two types of activity, active galactic nucleus (AGN) and starburst, 
are thought to contribute to the far-infrared emission of IRLGs.
The estimate of the relative energetic contributions of these 
kinds of activity is an important issue for understanding the 
nature of IRLGs.
Furthermore, since the bulk of the cosmic submillimeter background 
emission can be explained by IRLGs at high-redshift (Blain et al. 1999), 
understanding the energy sources of nearby IRLGs is closely 
related to the origin of the background emission.

Both AGN and starburst activity produces strong emission lines.
Since many emission lines, such as narrow emission lines of hydrogen 
or H$_{2}$ molecules, can be caused by both kinds of activity, 
it is not easy to distinguish their origin.
The polycyclic aromatic hydrocarbon (PAH) emission features can be a 
powerful tool for separating these kinds of activity and estimating 
their relative importance because the features are associated only with 
starburst activity and not with AGN activity (Moorwood 1986). 
Using the PAH emission feature at 7.7 $\mu$m, systematic studies of 
the energy sources of IRLGs have been 
reported based on the {\it Infrared Space Observatory} ({\it ISO}) 
spectra at 5.8--11.6 $\mu$m 
(Lutz et al. 1998; Genzel et al. 1998; Rigopoulou et al. 1999).
The observed ratios between the 7.7 $\mu$m PAH peak flux and 
7.7 $\mu$m continuum flux of IRLGs were estimated to be as high 
as those of starburst galaxies with lower far-infrared luminosities, 
leading these authors to suggest that 
starburst activity is a dominant energy source for most of the IRLGs 
they observed.
However, the 7.7 $\mu$m PAH to far-infrared luminosity ratios of IRLGs 
are roughly half of those of 
starburst galaxies (Genzel \& Cesarsky 2000). 
This indicates that observed starburst activity 
can account for only half of the far-infrared luminosities for IRLGs, 
although higher dust extinction toward starburst activity in IRLGs 
could make the intrinsic contribution of starburst activity higher.

In the 5.8--11.6 $\mu$m spectra of IRLGs, particularly obscured ones, 
not only the PAH emission, 
but also broad silicate dust absorption, is likely to exist 
at 8--13 $\mu$m (Mathis 1990).
Limited wavelength coverage of most {\it ISO} spectra (5.8--11.6 $\mu$m) 
does not allow the determination of a secure continuum level for 
many IRLGs and thus the estimate of the PAH flux could be 
quantitatively uncertain, as mentioned in Rigopoulou et al. (1999).
In fact, the 7.7 $\mu$m PAH emission flux of Arp 220, an IRLG 
for which an {\it ISO} spectrum with wider wavelength coverage 
(5--16 $\mu$m) is available, could be considerably smaller than 
that estimated by Rigopoulou et al., if the continuum level is determined 
in a different way (Elbaz et al. 1998; Dudley 1999).
Furthermore, in the case of another IRLG, IRAS 05189$-$2524, 
the reported equivalent widths of the 7.7 $\mu$m PAH emission 
feature based on the {\it ISO} spectrum differ by a factor of $>$3 
according to different authors (Clavel et al. 2000; Laureijs et al. 2000).

Spectroscopy in the range 3--4 $\mu$m can be a powerful method for 
investigating the energy source of IRLGs, by avoiding the 
above uncertainty.
First, since dust extinction at 3--4 $\mu$m is estimated to be 
as small as that at 7--8 $\mu$m 
($A_{3.5 \mu m}$/$A_{7.5 \mu m}$ $\sim$ 0.8; Lutz et al. 1996), 
we can detect signs of obscured energy sources.
Next, we can diagnose the relative importance of AGN and starburst 
activity using spectral features in the range 3--4 $\mu$m.
If an IRLG is powered by starburst activity, 
PAH emission should be detected at 3.3 $\mu$m and sometimes 
at 3.4 $\mu$m (Tokunaga et al. 1991).
If a source is powered by obscured AGN activity, 
a 3.4 $\mu$m absorption feature of carbonaceous dust grains 
should be detected (Pendleton et al. 1994), as found 
in some sources (e.g., NGC 1068; Bridger, Wright \& Geballe 1994; 
Imanishi et al. 1997).
If a source is powered by unobscured AGN activity, a 3--4 $\mu$m 
spectrum should be nearly featureless. 
Figure 1 summarizes the expected spectral shapes for the above 
three cases and for the cases of composites of AGN and 
starburst activity.
Finally, since a continuum level both longward and shortward of these 
features at 3.3--3.4 $\mu$m is observable in the $L$-band 
(2.8--4.2 $\mu$m) atmospheric window of Earth for nearby 
({\it z} $<$ 0.18) sources, there is no serious uncertainty 
in determining a continuum level.

Ground-based spectroscopy usually makes use of small slit widths, 
typically less than a few arcsec.
One arcsec corresponds to greater than a few 100 pc in IRLGs at 
z $>$ 0.02.
There is currently an increasing amount of evidence 
based on high spatial resolution imaging studies of infrared 
2--25 $\mu$m continuum emission that the bulk of 
the huge dust emission luminosities of IRLGs come from compact 
($<$ a few 100 pc) 
regions powered either by AGN and/or compact starburst activity, 
with little contribution from extended regions  
(Surace \& Sanders 1999; Surace, Sanders, \& Evans 2000; 
Scoville et al. 2000; Soifer et al. 2000).
Thus, ground-based 3--4 $\mu$m spectroscopy can fully 
diagnose the properties of dust emission from the energetically 
important compact regions in most IRLGs.
The present study reports ground-based spectroscopy in the 
wavelength range 3--4 $\mu$m of nine IRLGs to investigate 
their energy sources.

\section{Target Selection}

To achieve signal-to-noise ratios higher than 10--20 
with a few hours of on-source integration time using the 3.8-m 
United Kingdom Infrared Telescope (UKIRT), 
IRLGs brighter than 11 mag at $L$ or $L'$ were selected.
The targets are summarized in Table 1.
A total of nine IRLGs were observed.
Far-infrared luminosities are higher than $\sim$10$^{45}$ ergs s$^{-1}$.

In addition to the IRLGs, NGC 253, a nearby 
($\sim$3 Mpc; Tully 1988) galaxy, was also observed.
The emission properties of NGC 253 can be explained by starburst 
activity only (e.g., Rieke et al. 1980; Engelbracht et al. 1998), 
and no strong AGN activity has been found.
Although its far-infrared luminosity (Table 1) is more than an order 
of magnitude smaller than those of the IRLGs,  
this source is powered by compact ($\sim$100 pc) starburst activity 
(Kalas \& Wynn-Williams 1994). 
NGC 253 is thus an important nearby source for studying the nature 
of compact starburst activity found in IRLGs, 
and its 3--4 $\mu$m spectrum can be used as a template of 
compact starburst activity.

\section{Observation and Data Analysis}

We used the cooled grating spectrometer (CGS4; Mountain et al. 1990) 
to obtain 3--4 $\mu$m spectra of the IRLGs and NGC 253 with UKIRT 
on Mauna Kea, Hawaii.
An observing log is summarized in Table 2.
Sky conditions were photometric throughout the observations. 
The detector was a 256 $\times$ 256 InSb array.
The 40 l mm$^{-1}$ grating with 2 pixels wide slit (= 1$\farcs2$) 
was used.
The resulting spectral resolution was $\sim$750 at 3.5 $\mu$m.

Spectra were obtained toward the flux peak at 3--4 $\mu$m.
A standard telescope nodding technique was employed along the slit 
to subtract the signal from the sky.
The nodding amplitude was $\sim$12$''$ except for NGC 253, for which 
it was $\sim$24$''$.
The position angles of the slit used are summarized in Table 2.
For IRLGs that have clear double nuclei with a small separation 
($<$ several arcsec; Arp 220, NGC 6240, Mrk 273, and IRAS 08572+3915; 
Scoville et al. 2000), 
the position angles were set so that signals from both the nuclei 
were observed simultaneously. 

A- to G-type standard stars (Table 2) were observed with almost the same 
airmass as the target sources to correct for the transmission of 
Earth's atmosphere.
For target sources whose standard stars are A-type, 
data points at 3.74 $\mu$m and 4.05 $\mu$m were removed from final 
spectra because these data points are affected by, respectively, 
strong Pf$\gamma$ and Br$\alpha$ absorption lines in the spectra of 
A-type stars.
The $L$- or $L'$-band magnitudes of standard stars were estimated 
from their $V$-band magnitudes, by adopting the $V-L$ or $V-L'$ colors of 
the stellar types of individual standard stars (Tokunaga 2000).

Standard data analysis procedures were employed using IRAF.  
\footnote{
IRAF is distributed by the National Optical Astronomy Observatories, 
which are operated by the Association of Universities for Research 
in Astronomy, Inc. (AURA), under cooperative agreement with the 
National Science Foundation.} 
First, data of bad pixels were replaced with the interpolated signals 
of the surrounding pixels.
Next, bias was subtracted from the obtained frames and 
the frames were divided by a flat image.
Then, the spectra of the targets and the standard stars were extracted.
Since a secondary nucleus was not clearly recognizable as a 
distinct nucleus in all cases, spectra were extracted by integrating 
signals over 8$''$--20$''$ along the slit (Table 3).
Wavelength calibration was performed using an argon or a krypton lamp, 
and is believed to be accurate within 0.005 $\mu$m.
The galaxy spectra were divided by those of the standard stars, 
and were multiplied by the spectra of blackbodies with temperatures 
corresponding to individual standard stars (Table 2). 
After flux calibration based on the adopted standard star fluxes, 
the final spectra were produced.

\section{Results}

The flux-calibrated spectra are shown in Figure 2.
Spectra for some IRLGs were available in the literature, and 
the new spectra generally agree with previously published spectra. 
Detailed comparisons of our spectra with previous ones are found in 
Appendix A.

The comparison of the magnitudes of the CGS4 data with those in 
the literature is summarized in Table 3.
The magnitudes of IRLGs based on the CGS4 data roughly agree with 
photometric magnitudes measured with larger apertures, suggesting 
that nearly all of previously measured 3--4 $\mu$m emission is also 
detected in the CGS4 data.

The spectra of IC 694, Arp 220, NGC 6240, and Mrk 273 are dominated by 
the 3.3 $\mu$m PAH emission and no detectable 3.4 $\mu$m carbonaceous 
dust absorption, as in the case of NGC 253.
The spectrum of Mrk 231 shows the 3.3 $\mu$m PAH emission and 
possible 3.4 $\mu$m carbonaceous dust absorption, 
but the reality of the latter feature is currently doubtful.
The spectrum of IRAS 05189$-$2524 shows both 3.3 $\mu$m PAH emission 
and detectable 3.4 $\mu$m carbonaceous absorption 
with an optical depth of $\tau_{3.4}$ $\sim$ 0.04.
The spectrum of IRAS 08572+3915 is dominated by the 
3.4 $\mu$m carbonaceous dust absorption feature 
($\tau_{3.4}$ $\sim$ 0.9).
Neither detectable 3.3 $\mu$m PAH emission nor the 3.4 $\mu$m 
carbonaceous dust absorption is found in the spectra of 
IRAS 20460+1925 and IRAS 23060+0505. 
The fluxes and luminosities of the 3.3 $\mu$m PAH emission feature
are summarized in Table 4.

In addition to the above features, Pf$\gamma$ emission lines 
(rest wavelength = 3.74 $\mu$m) are clearly detected 
in the spectra of IC 694 and NGC 253.
The observed fluxes are 2 $\times$ 10$^{-14}$ ergs s$^{-1}$ cm$^{-2}$ 
and 6 $\times$ 10$^{-14}$ ergs s$^{-1}$ cm$^{-2}$ for 
IC 694 and NGC 253, respectively.
In the spectrum of NGC 6240, we attribute the flux excess at 
3.71 $\mu$m as H$_{2}$ 0--0 S(15) emission line 
(rest wavelength = 3.626 $\mu$m).
Its flux is 1 $\times$ 10$^{-14}$ ergs s$^{-1}$ cm$^{-2}$.  
NGC 6240 is known to show strong H$_{2}$ emission lines 
whose strength ratios can be approximated by thermal excitation 
with the temperature of $\sim$2000 K (Sugai et al. 1997).
The comparison with the H$_{2}$ 1--0 S(0) flux 
(1.8 $\times$ 10$^{-13}$ ergs s$^{-1}$ cm$^{-2}$) measured with 
a 3$''$ $\times$ 6$''$ aperture (Goldader et al. 1995) 
provides the 0--0 S(15) to 1--0 S(1) flux ratio of 0.06.
This ratio is similar to that found in shock gas in 
the Orion molecular outflow where the H$_{2}$ line ratios are roughly 
fitted by 2000 K thermal emission (Brand et al. 1988).

\section{Discussion} 

In IRLGs, AGN activity and compact starburst activity are likely to 
coexist spatially close to each other.
If energetic ionizing photons from AGN activity penetrate the compact 
starburst regions and destroy PAHs (Voit 1992) that otherwise 
should produce the strong 3.3 $\mu$m emission feature, 
the 3.3 $\mu$m PAH luminosity could be small in spite of the 
presence of strong compact starburst activity.
However, in IRLGs, the destruction of PAHs in the starburst regions by 
energetic photons from AGN activity probably does not occur because of 
the large column density and high pressures in the interstellar 
medium (Maloney 1999).
In the following discussion, it is assumed that compact starburst 
activity in the IRLGs produces the 3.3 $\mu$m PAH emission 
intrinsically as strongly as the star-formation activity 
in starburst galaxies with lower far-infrared luminosities, namely,  
in proportion to the far-infrared luminnosities.

\subsection{The Equivalent Width of the 3.3 $\mu$m PAH Emission Feature}

For starburst-dominated galaxies, the rest-frame equivalent widths of 
the 3.3 $\mu$m PAH emission feature (EW$_{\rm 3.3}$) do not 
change significantly as a result of dust extinction  
because flux attenuation of the 3.3 $\mu$m PAH emission and 3--4 $\mu$m 
continuum is similar.
Thus very small EW$_{3.3}$ values are most likely caused by 
the contribution of AGN activity to the 3--4 $\mu$m continuum emission.

The rest-frame EW$_{\rm 3.3}$ values of the nine IRLGs are summarized 
in Table 4.
The value for NGC 253 is 120 nm, which is nearly a median value 
for starburst-dominated galaxies (Moorwood 1986).
If the scatter of the values is taken into account and IRLGs with 
EW$_{\rm 3.3}$ = 60--240 nm are regarded as candidate 
starburst-dominated galaxies, then 
only IC 694, Arp 220, and NGC 6240 have EW$_{\rm 3.3}$ values 
as high as those of starburst galaxies, and are thus 
candidate starburst-dominated galaxies.
Mrk 273 shows a factor of 
${^{\displaystyle >}_{\displaystyle \sim}}$2 smaller value than 
those of starburst galaxies and thus AGN activity is estimated to 
contribute ${^{\displaystyle >}_{\displaystyle \sim}}$50\% of 
the 3--4 $\mu$m continuum flux.   
The remaining five IRLGs (Mrk 231, IRAS 05189$-$2524, IRAS 08572+3915, 
IRAS 23060+0505, and IRAS 20460+1925) have EW$_{\rm 3.3}$ ratios more 
than an order of magnitude smaller, indicating that $>$90\% of 
the 3--4 $\mu$m continuum flux originates in AGN activity.
Since the far-infrared to 3--4 $\mu$m flux ratio of dust emission 
heated by AGN activity through a radiative transfer mechanism  
depends strongly on dust radial density distribution 
(Ivezic, Nenkova \& Elitzur 1999), conversion from the AGN 
contribution at 3--4 $\mu$m to that at far-infrared wavelengths is 
not certain.
However, the EW$_{\rm 3.3}$ values for these six IRLGs 
(Mrk 273 and the five IRLGs) give no evidence that they are powered 
predominantly by compact starburst activity.

\subsection{The 3.3 $\mu$m PAH Emission to Far-Infrared 
Luminosity Ratios}

In $\S$5.1, three IRLGs have EW$_{\rm 3.3}$ values as high as those 
of starburst galaxies.
However, large EW$_{\rm 3.3}$ values of the IRLGs do not necessarily 
mean that their far-infrared luminosities are powered predominantly 
by starburst activity. 
Although highly obscured AGN activity can contribute to the observed 
far-infrared luminosity because of negligible extinction 
in this wavelength range, 
the contribution of the highly obscured AGN to the 3--4 $\mu$m flux 
can be highly attenuated.
As shown in Figure 1$e$, the apparent spectral shapes of galaxies 
powered both by starburst and highly obscured AGN activity 
could be nearly the same as those powered only by 
starburst activity (Fig. 1$a$).

A straightforward means of estimating the energy contribution of 
compact starburst activity is to estimate the absolute 3.3 $\mu$m PAH 
luminosity and compare it with the far-infrared luminosity.
The 3.3 $\mu$m PAH emission to far-infrared luminosity ratios 
($L_{3.3}$/$L_{\rm FIR}$) of the IRLGs are summarized in Table 4. 
The ratios of nearby starburst galaxies with lower far-infrared 
luminosity are estimated to be $\sim$1 $\times$ 10$^{-3}$ 
(Mouri et al. 1990), although the scatter in this ratio is difficult 
to estimate because, in the case of nearby starburst galaxies, 
the ratios are affected by different aperture sizes 
between 3--4 $\mu$m and far-infrared data (Mouri et al. 1990). 
The observed $L_{3.3}$/$L_{\rm FIR}$ ratios for all nine IRLGs 
are less than 3 $\times$ 10$^{-4}$, a factor of $>$3 smaller than 
those of starburst galaxies.
The smaller ratios for the six IRLGs with small EW$_{\rm 3.3}$ values 
can be explained naturally by a dominant contribution of AGN activity,  
while a few mechanisms can explain the smaller ratios in 
the three IRLGs with large EW$_{\rm 3.3}$ values 
(IC 694, Arp 220, and NGC 6240).

If highly obscured AGN activity contributes significantly 
to the far-infrared luminosity but little to the 3--4 $\mu$m flux 
due to high flux attenuation, then the large EW$_{\rm 3.3}$ values  
and the small $L_{3.3}$/$L_{\rm FIR}$ ratios can be explained.
This is the first possible mechanism.

A second mechanism is the effect of possible signal loss in 
the CGS4 data.
The possible signal loss for NGC 6240 could be 
as high as 0.5 mag or a factor of 1.6, while that for IC 694 and Arp 220 
is a factor of ${^{\displaystyle <}_{\displaystyle \sim}}$1.2.
Provided that spectral shape outside the CGS4 slit is the same as 
that inside it, the possible slit loss could decrease the 
$L_{3.3}$/$L_{\rm FIR}$ ratios only slightly.

A third possible mechanism is the effect of dust extinction.
The $L_{3.3}$/$L_{\rm FIR}$ ratios for starburst-dominated IRLGs 
could be smaller than those of starburst galaxies with lower 
far-infrared luminosities if dust extinction toward compact starburst 
activity in the IRLGs is much higher. 

For IC 694, Arp 220, and NGC 6240, dust extinction 
toward starburst regions in the case of a foreground dust screen 
model could be as high as $A_{\rm V}$ $\sim$ 20 mag 
(Sugai et al. 1999; Alonso-Herrero et al. 2000),  
$A_{\rm V}$ $\sim$ 45 mag (Rigopoulou et al. 1999, Fig. 7), 
and $A_{\rm V}$ $\sim$ 7 mag ($A_{\rm K}$ = 0.8 mag; Sugai et al. 1997), 
respectively. 
By adopting a standard extinction curve 
($A_{\rm L}$ $\sim$ 0.06--0.07 $\times$ $A_{\rm V}$; Rieke \& Lebofsky 
1985; Lutz et al. 1996) and taking possible signal loss into account, 
the {\it intrinsic} $L_{3.3}$/$L_{\rm FIR}$ ratios 
after correction for dust extinction could be 
as high as 4--7 $\times$ 10$^{-4}$ for all three.
Provided that the $L_{3.3}$/$L_{\rm FIR}$ ratios of control 
starburst galaxy samples suffer the effect of dust extinction 
with $A_{\rm V}$ $\sim$ 10 mag (Rigopoulou et al. 1999, Fig. 7), 
the intrinsic $L_{3.3}$/$L_{\rm FIR}$ ratios of 
starburst-dominated galaxies are estimated to be 
1.7--1.9 $\times$ 10$^{-3}$.
The intrinsic $L_{3.3}$/$L_{\rm FIR}$ ratios for the three 
IRLGs are a factor of $\sim$3 smaller.
For reference, the 3.3 $\mu$m PAH flux of NGC 253 within our slit 
is measured to be 2.8 $\times$ 10$^{-12}$ ergs s$^{-1}$ cm$^{-2}$.
The $L$-band magnitude inside our slit is 
2.7 mag fainter than the total magnitude of this galaxy (Table 3).
If the 3--4 $\mu$m spectrum shape outside the slit is 
the same as that inside it, and if $A_{\rm V}$ = 14 mag is adopted 
as dust extinction toward starburst regions in this galaxy 
(Rigopoulou et al. 1999, Fig. 7), then the intrinsic 
$L_{3.3}$/$L_{\rm FIR}$ ratio is 1.3--1.5 $\times$ 10$^{-3}$, 
consistent with the value of 1.7--1.9 $\times$ 10$^{-3}$ 
within $\sim$25\%. 

Although dust extinction for Arp 220 was derived based on mid-infrared 
data, the dust extinction for IC 694 and NGC 6240 was estimated 
based on near-infrared data. 
The dust extinction estimate based on near-infrared data could be smaller 
than that based on mid-infrared data; the latter estimate is thought 
to be more reliable, particularly in the case of high dust extinction 
(Genzel et al. 1998).  
From this viewpoint, it is possible that the intrinsic 
$L_{3.3}$/$L_{\rm FIR}$ ratios for IC 694 and NGC 6240 could be higher 
than our estimate.
Further, when dust is mixed with emission sources, dust extinction 
correction factor could be higher than that for a foreground screen 
dust model (e.g., Genzel et al. 1995).
Since the scatter of the intrinsic $L_{3.3}$/$L_{\rm FIR}$ ratios 
for starburst galaxies is also poorly known, the energy source of 
these three IRLGs cannot be determined confidently from our diagnoses.

\subsection{Summary of the Energy Sources Derived from the 3--4 $\mu$m 
Spectra}

The energy sources of the nine IRLGs derived from the 3--4 $\mu$m 
spectra is summarized in Table 5.
For IC 694, Arp 220, and NGC 6240, which show strong 3.3 $\mu$m PAH 
emission in equivalent width, it is not clear whether 
compact starburst activity is the only important energy source or 
a significant fraction of the luminosity is powered by obscured 
AGN activity (Figs. 1$a$ and 1$e$).
Mrk 231 and Mrk 273, which show weak 3.3 $\mu$m PAH 
emission and no detectable 3.4 $\mu$m carbonaceous dust absorption,  
are suggested to be powered by either unobscured or obscured AGN activity 
(Figs. 1$d$ and 1$e$).
IRAS 05189$-$2524, which shows weak 3.3 $\mu$m PAH emission and 
detectable (but weak) 3.4 $\mu$m carbonaceous dust absorption,  
is thought to be powered by moderately obscured AGN activity (Fig. 1$e$).
The 3--4 $\mu$m spectrum of IRAS 08572+3915 is typical of 
highly dust obscured AGNs, and thus this source is most likely 
powered by highly embedded AGN activity (Fig. 1$b$), as 
further supported by more detail study based on 3--20 $\mu$m 
spectroscopy (Dudley \& Wynn-Williams 1997; Imanishi \& Ueno 2000). 
The featureless 3--4 $\mu$m spectra of IRAS 23060+0505 and 
IRAS 20460+1925 suggest that these sources are powered by 
less obscured AGN activity (Fig. 1$c$).

\subsection{Comparison with Other Diagnoses}

\subsubsection{The $ISO$ Spectra}

For Arp 220, NGC 6240, Mrk 273, Mrk 231, IRAS 05189$-$2524, and 
IRAS 23060+0505, the 7.7 $\mu$m PAH peak flux to 7.7 $\mu$m continuum 
flux ratios were used to diagnose their energy source 
(Rigopoulou et al. 1999; Laureijs et al. 2000).
These ratios and the equivalent widths are a similar (but not identical) 
concept, and both provide the PAH strength relative to the continuum 
emission. 
The results of the 3--4 $\mu$m spectra and the {\it ISO} spectra 
are compared in Table 5.
As mentioned in $\S$ 1, a continuum determination based on the 
{\it ISO} 5.8--11.6 $\mu$m spectra is not easy. 
When data at 11 $\mu$m were used to determine a continuum level 
(Rigopoulou et al. 1999), the 7.7 $\mu$m PAH flux for sources 
with a strong silicate dust absorption feature could be overestimated.
Since a different amount of effect of dust extinction at different 
wavelengths does not change the relative PAH strengths significantly 
($\S$ 5.1), 
the comparison of the strength of the 3.3 $\mu$m and 7.7 $\mu$m 
PAH emission relative to the continuum could provide useful information 
on the effect of the silicate dust absorption in the {\it ISO} spectra.

The relative PAH strengths (EW$_{\rm 3.3}$ and 7.7 $\mu$m PAH peak to 
continuum flux ratio) for Mrk 231, IRAS 05189$-$2524, and 
IRAS 23060+0505 are much smaller than those of starburst galaxies 
in both wavelengths.
The relative PAH strength for NGC 6240 is the same within 30\% as the 
typical value of starburst-dominated galaxies at both wavelengths.
In fact, for NGC 6240, no indication of strong silicate dust absorption 
was found in ground-based 8--13 $\mu$m spectra 
(Roche et al. 1991; Dudley 1999).
However, for Arp 220,  
the 3--4 $\mu$m spectrum provides 
the EW$_{\rm 3.3}$ value of $\sim$80 nm, slightly smaller 
than those of starburst-dominated galaxies ($\sim$120 nm), 
while the {\it ISO} spectrum provides the 7.7 $\mu$m PAH 
peak-to-continuum flux ratio of $\sim$4.2, significantly higher than 
the average value of starburst galaxies ($\sim$3.0). 
Also for Mrk 273, the 3--4 $\mu$m spectrum gives 
EW$_{\rm 3.3}$ $\sim$35 nm, a factor of $\sim$3 smaller than 
starburst galaxies, while the {\it ISO} spectrum gives the 7.7 $\mu$m 
PAH peak-to-continuum flux ratio $\sim$1.9, only $\sim$40\% smaller.
For Arp 220 and Mrk 273, Dudley (1999) argued that 
silicate dust absorption plays an important role based on 
ground-based 8--13 $\mu$m spectra.
The discrepancy may be caused by the effect of the silicate dust 
absorption.

The above comparison suggests that, for at least a some fraction 
of IRLGs, the 7.7 $\mu$m PAH flux estimated by Rigopoulou et al. (1999) 
could be too large because of the use of data affected by silicate 
dust absorption as a continuum level.
In this case, the presence of the silicate dust absorption is 
underestimated. 
Highly obscured AGNs show spectra characterized by weak PAH emission 
and strong silicate dust absorption.
Virtually no IRLGs powered by highly obscured AGNs were found 
based on the {\it ISO} data (Genzel et al. 1998; Rigopoulou et al. 1999), 
which may be the result of the determination of the continuum level.

Finally, for IRAS 05189$-$2524, the flux and equivalent 
width of the 3.3 $\mu$m PAH emission estimated by an {\it ISO} spectrum 
obtained using a larger (24$''$ $\times$ 24$''$) aperture 
(Clavel et al. 2000) are a factor of $\sim$6 larger than the estimate 
in this paper.
If the larger values are the case, the bulk of the 3.3 $\mu$m PAH 
emission detected in the {\it ISO} spectrum must originate 
in extended star-forming regions outside the CGS4 slit.
The extended star-formation regions could explain nearly all of 
the far-infrared luminosity of this source, if the value of 
$L_{3.3}$/$L_{\rm FIR}$ = 1.7--1.9 $\times$ 10$^{-3}$ is adopted.
However, this source shows no extended continuum emission at 2--25 $\mu$m 
(Scoville et al. 2000; Soifer et al. 2000).
If the extended regions are fully responsible for the far-infrared 
emission at 60 $\mu$m but contribute to, say, less than 5\% of 
the 25 $\mu$m flux (Soifer et al. 2000), then 
the 60 $\mu$m to 25 $\mu$m flux ratio in Jy is $>$80, 
much larger than any nearby star-forming galaxies (Rice et al. 1988).
Since the physical conditions in the extended regions of 
IRAS 05189$-$2524 (not nuclear compact regions that could have 
an extreme environment in density and so on) are expected not to be 
quite different from those in nearby star-forming galaxies,  
such a sudden flux increase from the mid-infrared ($<$25 $\mu$m) 
to far-infrared ($>$40 $\mu$m) only in the case of 
IRAS 05189$-$2524 is very unlikely.
While the 3.3 $\mu$m PAH flux in this paper is estimated by tracing 
the spectral profile (see the caption of 
Table 4 for more details), Clavel et al. simply sum up  
all the flux above the adopted continuum level in the pre-defined 
wavelength range (J. Clavel, private communication).
The {\it ISO} spectrum provided by Dr. Clavel (private communication) 
shows a flux excess above an adopted continuum level at $\sim$3.3 $\mu$m, 
but it has lower spectral resolution (R $\sim$ 100) 
and larger scatter in data points at 3--4 $\mu$m than the CGS4 spectrum.
Furthermore, although the wavelength of the 3.3 $\mu$m PAH emission 
feature usually extends from 3.24 $\mu$m to 3.34 $\mu$m in the 
rest-frame (Tokunaga et al. 1991; this work), 
the flux excess in the {\it ISO} spectrum is found from 
3.25 $\mu$m to 3.40 $\mu$m (not 3.34 $\mu$m).
If this excess is actually due to the 3.3 $\mu$m PAH emission feature, 
its spectral shape in the extended regions of IRAS 05189$-$2524 
is quite different from typical ones.
More sensitive and higher spectral resolution spectroscopy with a wide 
aperture is necessary to address this issue.
We tentatively consider that the extended regions contribute little to 
the total far-infrared luminosity of IRAS 05189$-$2524.

\subsubsection{8--25 $\mu$m Imaging Studies}

Soifer et al. (2000) suggested based on their 8--25 $\mu$m imaging 
data that AGN activity is the dominant energy 
source of Mrk 231, IRAS 05189$-$2524, and IRAS 08572+3915.
Our suggestions are the same for these three sources.
Soifer et al. were unable to distinguish the energy source of 
Mrk 273 and Arp 220.
We argue that AGN activity is the main energy source of Mrk 273, 
while the energy source of Arp 220 is not clear.
As a whole, the suggestions in this paper are consistent with those 
by Soifer et al.

\subsubsection{2 $\mu$m Spectroscopy}

In the $<$2 $\mu$m spectra of Mrk 231, IRAS 05189$-$2524, 
IRAS 23060+0505, and IRAS 20460+1925, a broad 
($>$2000 km s$^{-1}$ in full width at half-maximum) component of 
neutral hydrogen recombination line emission was detected, and 
their intrinsic 
luminosities are high enough to account for the far-infrared 
luminosities resulting from AGN activity alone 
(Veilleux, Sanders, \& Kim 1999b).
Our suggestions about the energy source for these four IRLGs are 
consistent with theirs.

\subsubsection{Summary of the Above Comparison}

Among the nine IRLGs, five (Mrk 231, IRAS 05189$-$2524, IRAS 08572+3915, 
IRAS 23060+0505, and IRAS 20460+1925) have 
warm ({\it IRAS} 25 $\mu$m to 60 $\mu$m flux 
ratio $f_{\rm 25}$/$f_{\rm 60}$ $>$ 0.2) far-infrared color (Table 1). 
The above three infrared diagnoses 
have provided a growing amount of evidence that these warm 
IRLGs are powered predominantly by AGN activity (Table 5).
However, warm IRLGs constitute only $\sim$20\% of total IRLGs,  
and most IRLGs have cool ($f_{\rm 25}$/$f_{\rm 60}$ $<$ 0.2) 
far-infrared color (Kim \& Sanders 1998). 
It was argued based on the estimate of the 7.7 $\mu$m PAH strengths 
in the {\it ISO} spectra that these cool IRLGs are powered predominantly 
by starburst activity, with little AGN contribution 
(Genzel \& Cesarsky 2000).
Among the nine IRLGs, three (Arp 220, NGC 6240, and Mrk 273) are 
cool IRLGs and are classified as starburst-dominated IRLGs based on the 
7.7 $\mu$m PAH strength (Table 5).
However, for Arp 220 and Mrk 273, our 3--4 $\mu$m spectra suggest that 
the 7.7 $\mu$m PAH strengths, and thereby the magnitudes of starburst 
activity, may be overestimated based on the {\it ISO} spectra. 
In fact, for Mrk 273, although the relative 7.7 $\mu$m PAH 
emission strength suggests this source is powered by starburst 
activity (Rigopoulou et al. 1999), a different diagnosis based on 
high excitation 
lines in the {\it ISO} spectra indicates that AGN activity is a 
dominant energy source (Genzel et al. 1998).
For NGC 6240, X-ray data suggest that highly embedded and 
highly luminous AGN activity is an important energy source 
(Vignati et al. 1999).   
The contribution of AGN activity in the cool IRLGs may be higher 
than the estimate based on the 7.7 $\mu$m PAH emission strength in 
the {\it ISO} spectra. 

\subsubsection{Compact Radio Core Emission}

High spatial resolution radio images have been used to investigate the 
relative importance between AGN and starburst activity in IRLGs.
A compact radio core, which was thought to be strong 
evidence for AGN activity, is no longer definitive evidence for 
AGN activity because both AGN and compact starburst activity 
could explain observed properties of the compact radio core 
(Smith, Lonsdale, \& Lonsdale 1998). 
However, IRLGs whose radio emission morphology is extended without 
a compact core are thought to be powered by starburst activity.
Compact radio cores were detected in IC 694, Arp 220, NGC 6240, 
Mrk 273, Mrk 231, and IRAS 20460+1925 
(Lonsdale, Lonsdale, \& Smith 1992; Lonsdale, Smith, \& Lonsdale 1993; 
Heisler et al. 1998; Kewley et al. 2000), but not in 
IRAS 05189+2524, IRAS 08572+3915, and IRAS 23060+0505 
(Lonsdale et al. 1993; Heisler et al. 1998). 
It is puzzling that no radio core emission was detected 
in the three warm IRLGs, for which more than one infrared diagnoses 
consistently suggest that AGN activity is a dominant energy source  
(Table 5).

\subsection{Future Study}

A high fraction (6/9) of IRLGs in this sample are suggested to be 
powered by AGN activity.
This is thought to be caused by the selection effect of picking 
IRLGs that are bright at 3--4 $\mu$m.
At a given far-infrared flux, the 3--4 $\mu$m flux of less obscured 
AGNs is generally higher than that of starburst-dominated galaxies 
(Schmitt et al. 1997).
Among the nine IRLGs, 5/9 (55\%) are optically classified as Seyfert 
types (Table 1) and 
4/9 (44\%; Mrk 231, IRAS 05189$-$2524, IRAS 23060+1925, and 
IRAS 23060+0505) show a detectable broad emission line at $<$2 $\mu$m 
(that is, AGN activity is less obscured).
These fractions are significantly higher than those  
in a larger sample of IRLGs ($\sim$20\%; Veilleux, Kim, \& Sanders 1999a; 
Veilleux et al. 1999b).
Furthermore, the fraction of warm IRLGs in this sample (5/9; 55\%) 
is much higher than that in a larger sample of IRLGs 
($\sim$20\%; Kim \& Sanders 1998).
Observations of a larger number of fainter IRLGs at $L$ are clearly 
necessary to evaluate the relative role between AGN and starburst 
activity  in IRLGs as a whole.

\section{Summary}  

The energy sources of nine IRLGs 
were diagnosed using the EW$_{\rm 3.3}$ values and 
the $L_{3.3}$/$L_{\rm FIR}$ ratios.
The following main results were found:

\begin{enumerate}
\item Mrk 231, Mrk 273, and IRAS 05189$-$2524 show detectable 
      3.3 $\mu$m PAH emission in their 3--4 $\mu$m spectra. 
      However, the PAH emission strengths are too small to 
      account for the bulk of their far-infrared luminosities by 
      starburst activity,  
      indicating that these sources are powered by AGN activity.
\item No PAH emission was detected in the spectra of 
      IRAS 08572+3915, IRAS 23060+0505, and IRAS 20460+1925.  
      It was suggested that IRAS 08572+3915 is powered by 
      highly obscured AGN activity, while IRAS 23060+0505 and 
      IRAS 20460+1925 are powered by less-obscured AGN activity.
\item For IC 694, Arp 220, and NGC 6240, 
      while the EW$_{\rm 3.3}$ values are as large as those of 
      starburst galaxies, the $L_{3.3}$/$L_{\rm FIR}$ ratios 
      after correction for screen dust extinction are a factor of 
      $\sim$3 smaller.
      Whether these three IRLGs are powered by starburst activity only 
      or obscured AGN activity contributes significantly is not clear 
      because actual dust extinction correction factor could be higher 
      than our estimate and the scatter of the intrinsic 
      $L_{3.3}$/$L_{\rm FIR}$ ratios for starburst activity is poorly 
      known.
\item Our energy diagnoses reinforced the previous suggestions that 
      AGN activity is the dominant energy source in warm IRLGs.
\end{enumerate}

\acknowledgments      

We thank Dr. T. Kerr, Dr. J. Davies, T. Carroll, and T. Wold for their 
support during the UKIRT observing runs.
We are grateful to Dr. D. B. Sanders and the anonymous referee for 
their useful comments.
Drs. A. T. Tokunaga and H. Ando give MI the opportunity to work at 
the University of Hawaii.
L. Good kindly proofread this paper, and K. Teramura modified Fig. 1.
The United Kingdom Infrared Telescope is operated by the 
Joint Astronomy Centre on behalf of the U.K. Particle Physics and 
Astronomy Research Council.
MI is financially supported by the Japan Society for the Promotion 
of Science during his stay at the University of Hawaii.
CCD gratefully acknowledges the National Research Council (USA), the 
Naval Research Laboratory, and the Office of Naval Research for their
generous support, and the sponsorship of Dr. J. Fischer.

\clearpage 
 
\appendix
\section{Comparisons with Previous 3--4 $\mu$m Spectra}

Our new spectra (mostly higher signal-to-noise ratios and spectral 
resolution) are compared with old spectra in the literature.

\noindent {\it IC 694} 
\indent 

A flux-calibrated 3--4 $\mu$m spectrum taken with a 2$\farcs$7 aperture 
was presented by Dennefeld \& Desert (1990).
The overall spectral shape and flux level in the new spectrum are 
consistent with the previous spectrum.
Using narrow band filters, Satyapal et al. (1999) estimated the flux 
of the 3.3 $\mu$m PAH emission to be 
8.5 $\times$ 10$^{-13}$ ergs s$^{-1}$ cm$^{-2}$ 
within a 7$''$ aperture. 
This is $\sim$40\% higher than our estimate, but this difference does 
not affect our discussion significantly.

\noindent {\it Arp 220 and NGC 6240}
\indent 

Low-resolution spectra taken using a circularly variable filter 
with a 8$\farcs$7 aperture were presented by Rieke et al. (1985).
They showed strong 3.3 $\mu$m PAH emission.
The new spectra show slightly smaller continuum flux levels but similar 
spectral shapes.

\noindent {\it Mrk 231}
\indent 

A flux-calibrated 3--4 $\mu$m spectrum taken with a 3$\farcs$8 aperture 
was presented by Imanishi et al. (1998). 
The new spectrum provides much higher signal-to-noise ratios.
The spectral shape and flux level are consistent with the previous 
spectrum. 

\noindent {\it IRAS 05189$-$2524}
\indent 

A flux-uncalibrated 3--4 $\mu$m spectrum was presented by Wright et al. 
(1996).
It showed weak 3.3 $\mu$m PAH emission and weak 3.4 $\mu$m carbonaceous 
dust absorption with an optical depth of $\sim$0.05.
The new spectrum is consistent with the previous spectrum in the spectral 
shape and the optical depth of the carbonaceous dust absorption.

\noindent {\it IRAS 08572+3915}
\indent 

A flux-uncalibrated 3--4 $\mu$m spectrum was presented by Wright et al. 
(1996).
It was dominated by the 3.4 $\mu$m carbonaceous dust absorption feature,  
and its optical depth was estimated to be $\sim$0.9 (Pendleton 1996).
Both the spectral shape and the optical depth in the new spectrum 
are consistent with the previous spectrum.

\noindent {\it IRAS 23060+0505}
\indent 

Wright et al. (1996) classified the 3--4 $\mu$m spectrum of 
IRAS 23060+0505 as featureless.
The new spectrum is consistent with their classification.

\noindent {\it NGC 253}
\indent 

A low-resolution spectrum taken using a circularly variable filter 
with a 7$\farcs$5 aperture was presented by Moorwood (1986).
Both the new and old spectra show the PAH emission at 3.3 $\mu$m and 
3.4 $\mu$m. 
The flux level in our new spectrum is smaller than the old spectrum 
because of smaller aperture size.

\clearpage

\small


\begin{center}
\begin{table}[t]
\caption{Nine IRLGs and NGC 253.}
\end{table}
\begin{tabular}{llcccccc} \hline \hline
       &     & $f_{25}$ & $f_{60}$ & $f_{100}$ & $L_{\rm FIR}$  
& Spectral & Infrared \\
Object & $z$ & (Jy) & (Jy) & (Jy) & (ergs s$^{-1}$) & Type   & Color \\ 
 (1)   & (2) & (3) & (4) & (5) & (6) & (7) & (8) \\ 
\hline
IC 694 (= Arp 299A) & 0.010 & 21.51$^{a}$ & 105.82 $^{a}$ & 
111.16 $^{a}$ & 44.90 $^{b}$ & Starburst (1) & ---  \\
Arp 220 (= IC 4553) & 0.018 & 8.11 & 104.08 & 117.69 & 45.63 & LINER (2) 
& Cool \\
NGC 6240 & 0.024 & 3.51 & 23.47 & 26.55 & 45.24 & LINER (3) & Cool \\
Mrk 273  & 0.038 & 2.33 & 23.70 & 22.31 & 45.62 & Seyfert 2 (2) & Cool \\
Mrk 231  & 0.042 & 8.52 & 33.60 & 30.89 & 45.86 & Seyfert 1 (2) &  Warm \\
IRAS 05189$-$2524 & 0.043 & 3.52 & 13.94 & 11.68 & 45.49 & Seyfert 2 (2) 
& Warm \\
IRAS 08572+3915 & 0.058 & 1.73 & 7.53 & 4.59 & 45.45 & LINER (2) & Warm \\
IRAS 23060+0505 & 0.174 & 0.46 & 1.15 & 0.83 & 45.63 & Seyfert 2 (2) 
& Warm \\
IRAS 20460+1925   & 0.181 & 0.54 & 0.88 & 0.48 & 45.52 & Seyfert 2 (4) 
& Warm \\ 
NGC 253           & 3 (Mpc) & 117.08 & 758.69 & 1044.66 & 43.76 
& Starburst $^{c}$ (5) & Cool \\ 
\hline
\end{tabular}
\end{center}

Note. --- Col. (1): Object name.
Col. (2): Redshift. For NGC 253, a very nearby galaxy, distance in Mpc 
            is shown.
Col. (3): $IRAS$ 25 $\mu$m flux in Jy.
Col. (4): $IRAS$ 60 $\mu$m flux in Jy.
Col. (5): $IRAS$ 100 $\mu$m flux in Jy.
Col. (6): Logarithm of far-infrared (40--500 $\mu$m) luminosity 
            in ergs s$^{-1}$ calculated with
            $L_{\rm FIR} = 2.1 \times 10^{39} \times$ $D$(Mpc)$^{2}$
            $\times (2.58 \times f_{60} + f_{100}$) ergs s$^{-1}$
            (Sanders \& Mirabel 1996).
Col. (7): Optical spectral type and reference in parentheses.  
            (1) Armus, Heckman, \& Miley 1989; 
            (2) Veilleux et al. 1999a; 
            (3) Veilleux et al. 1995; (4) Frogel et al. 1989;  
            (5) Engelbracht et al. 1998.
Col. (8): Far-infrared color.
            Warm ($f_{25}$/$f_{60}$ $>$ 0.2) or 
            Cool ($f_{25}$/$f_{60}$ $<$ 0.2). 

$^{a}$ Total flux of the multinuclei merging system Arp 299.
        IC 694 is one nucleus of Arp 299.
        The far-infrared emission of Arp 299 was spatially unresolved 
        with {\it IRAS}. 

$^{b}$ The airborne observations of Joy et al. (1989) showed that 
       about 60\% of the far-infrared luminosity of Arp 299 comes from 
       the IC 694 nucleus.

$^{c}$ Based on a near-infrared spectrum and modeling.

\newpage

\begin{center}
\begin{table}[t]
\caption{Observing Log.}
\end{table}
\begin{tabular}{lccccccc} \hline \hline
 & Date & Integration &  & 
\multicolumn{4}{c}{Standard Stars}  \\ 
Object & (UT) & Time (sec)& P.A. & Star Name & $L$-mag & Type & $T_{\rm eff}$ (K) \\ 
 (1)   & (2) & (3) & (4) & (5) & (6) & (7) & (8) \\ 
\hline
IC 694 & 2000 Feb 20--21 & 2560 & 0$^{\circ}$ &
HR 4761 & 4.8 & F6--8V & 6200 \\
Arp 220 & 2000 Feb 20 & 3840 & 90$^{\circ}$ & 
HR 5634 & 3.8 & F5V & 6500 \\
NGC 6240          & 1999 Sep 9 & 4800 & $-$19$^{\circ}$ &
HR 6629 & 3.8 & A0V & 9480 \\
Mrk 273 & 2000 Feb 20 & 2560 & $-$35$^{\circ}$ &
HR 4761 & 4.8 & F6--8V & 6200 \\
Mrk 231 & 2000 Feb 20 & 1280 & 0$^{\circ}$ & 
HR 4761 & 4.8 & F6--8V & 6200 \\
IRAS 05189$-$2524 & 1999 Sep 9 & 3200 & $-$19$^{\circ}$ &
HR 1762 & 4.7 & A0V & 9480 \\
IRAS 08572+3915   & 2000 Feb 20 & 2560 & 30$^{\circ}$ & 
HR 3625 & 4.5 & F9V & 6000 \\
IRAS 23060+0505   & 1999 Sep 8 & 5600 & 0$^{\circ}$ &
HR 8514 & 4.9 & F6V & 6400 \\ 
IRAS 20460+1925   & 1999 Sep 8 & 5600 & 0$^{\circ}$ & 
HR 7235 & 3.0 & A0V & 9480 \\
NGC 253           & 1999 Sep 9 & 2400 & 0$^{\circ}$ &
HR 448 & 4.3 & G2IV & 5830 \\ 
\hline
\end{tabular}
\end{center}

Note. --- Col. (1): Object name.
Col. (2): Observing date in UT.
Col. (3): Net on-source integration time in seconds.
Col. (4): Position angle of the slit.
            0$^{\circ}$ corresponds to the north-south direction.
            Position angle increases with clockwise on the sky plane. 
            For double nuclei sources with a small separation 
            (Arp 220, NGC 6240, Mrk 273, and IRAS 08572+3915), 
            the position angles were set in such a way that signals 
            from both the double nuclei enter the slit.
Cols. (5)-(8): Standard stars used to correct for the transmission of 
            Earth's atmosphere and to calibrate flux level.
Col. (5): Star name. 
Col. (6): $L$-band magnitude.
Col. (7): Spectral type.
Col. (8): Effective temperature.

\clearpage

\begin{table}[h]
\caption{$L$- or $L'$-band Magnitudes from CGS4 Data 
Compared with Magnitudes in the Literature.}
\end{table}

\begin{center}
\begin{tabular}{lccc} \hline \hline
       &   magnitudes & magnitudes  & Reference  \\
Object &  (CGS4)$^{a}$ & (literature)$^{b}$ &            \\ \hline
IC 694            & $L$=10.2 (1$\farcs$2 $\times$ 10$''$) & 
$L$=10.1 (4$''$)  & 1 \\
Arp 220           & $L$=10.7 (1$\farcs$2 $\times$ 10$''$) & 
$L$=10.5 (5$''$)  & 1\\
NGC 6240          & $L$=10.1 (1$\farcs$2 $\times$ 10$''$) & 
$L$=9.6 (9$''$)   & 2, 3 \\
Mrk 273           & $L$=10.5 (1$\farcs$2 $\times$ 10$''$) & 
$L$=10.5 (5$''$)  & 1 \\
Mrk 231           & $L$=7.2  (1$\farcs$2 $\times$ 8$''$) & 
$L$=7.4 (5$''$)   & 1 \\
IRAS 05189$-$2524 & $L$=8.6  (1$\farcs$2 $\times$ 8$''$) & 
$L$=8.4 (8$''$)   & 4 \\
IRAS 08572+3915   & $L$=10.1 (1$\farcs$2 $\times$ 8$''$) & 
$L$=10.0 (3$''$)  & 1 \\ 
IRAS 23060+0505   & $L'$=9.2 (1$\farcs$2 $\times$ 8$''$) & 
$L'$=9.1 (6$''$) & 5 \\  
IRAS 20460+1925   & $L'$=9.5 (1$\farcs$2 $\times$ 8$''$) & 
$L'$=?, $L$=9.2 (12$''$)  &  4, 6 \\
NGC 253           & $L$=8.4 (1$\farcs$2 $\times$ 20$''$) & 
$L$=5.7 (51$''$)   &  3 \\ 
\hline
\end{tabular}
\end{center}

$^{a}$ Derived from spectro-photometric CGS4 data taken with a 
1$\farcs$2 wide slit.

$^{b}$ Photometric data from the literature.

References.---
(1) Zhou, Wynn-Williams, \& Sanders 1993;  
(2) Rudy, Levan, \& Rodriguez-Espinosa 1982; 
(3) Glass \& Moorwood 1985; 
(4) Vader et al. 1993; 
(5) Hill, Wynn-Williams, \& Becklin 1987;
(6) Frogel et al. 1989.

\clearpage

\begin{center}
\begin{table}[t]
\caption{The Properties of the 3.3 $\mu$m PAH Emission Feature.}
\end{table}
\begin{tabular}{lcccc} \hline \hline
        &   log $f_{3.3}$ & log $L_{3.3}$ & & rest EW$_{3.3}$ \\
Object  & (ergs s$^{-1}$ cm$^{-2}$) & (ergs s$^{-1}$) & 
log $L_{3.3}$/$L_{\rm FIR}$ & (nm) \\ 
  (1)  & (2) & (3) & (4) & (5) \\
\hline
IC 694             & $-$12.23 & 41.05 & $-$3.85 & 150 \\
Arp 220 & $-$12.61 &  41.19 & $-$4.44 & 82  \\
NGC 6240           & $-$12.40 & 41.65 & $-$3.59  & 70 \\
Mrk 273     & $-$12.84 & 41.61 & $-$4.01 & 35  \\
Mrk 231 & $-$12.75 & 41.78 & $-$4.08 & 2.0  \\
IRAS 05189$-$2524  & $-$12.96 & 41.60 & $-$3.89 & 4.3 \\
IRAS 08572+3915 & $<$ $-$13.87 & $<$ 40.95 & $<$ $-$4.50 & $<$ 2.0  \\
IRAS 23060+0505    & $<$ $-$13.88 & $<$ 41.92 & $<$ $-$3.71 
& $<$ 1.0  \\ 
IRAS 20460+1925    & $<$ $-$14.30 & $<$ 41.53 
& $<$ $-$3.99 & $<$ 0.53 \\
NGC 253 & $-$11.56 & 39.47 & $-$4.29 & 120 \\
\hline
\end{tabular}
\end{center}

Note. --- Col. (1): Object name.
Col. (2): Logarithm of the observed flux of the 3.3 $\mu$m PAH emission 
            feature in ergs s$^{-1}$ cm$^{-2}$.
            The flux is estimated by assuming a Gaussian profile 
            whose rest-frame peak wavelength is 3.28--3.30 $\mu$m and 
            rest-frame line width in full width at half-maximum is 
            0.04--0.05 $\mu$m, as found in the starburst galaxies, 
            M82 (Tokunaga et al. 1991) and NGC 253 (this paper).
            The Gaussian profile can fit the 3.3 $\mu$m PAH emission 
            feature reasonably well for all the sources, and 
            systematic errors in the flux estimate using this method 
            are believed to be no more than $\sim$30\%.
Col. (3): Logarithm of the observed luminosity of the 3.3 $\mu$m PAH 
            emission feature in ergs s$^{-1}$.
Col. (4): Logarithm of the observed 3.3 $\mu$m PAH emission to 
            far-infrared luminosity ratio.
            It is about $-$3.00 for starburst-dominated galaxies.
Col. (5): Rest-frame equivalent width of the 3.3 $\mu$m PAH emission 
            feature in nm. 

\newpage

\begin{center}
\begin{table}[t]
\caption{Summary of the Main Energy Sources for the Nine IRLGs.}
\end{table}
\begin{tabular}{lcccc} \hline \hline
  (1)  & (2) & (3) & (4) & (5) \\
       & 3--4 $\mu$m & 5.8--11.6 $\mu$m & 8--25 $\mu$m & 2 $\mu$m \\ 
Object & Spectroscopy & {\it ISO} Spectroscopy & Imaging & Spectroscopy  \\ 
\hline
IC 694    & SB or SB+AGN   & --- & --- & --- \\
Arp 220   & SB or SB+AGN   & SB (4.2) &  ?  & --- \\
NGC 6240  & SB or SB+AGN   & SB (2.6) & --- & --- \\
Mrk 273              & AGN & SB (1.9) &  ?  & --- \\
Mrk 231              & AGN & AGN (0.3) & AGN & AGN \\
IRAS 05189$-$2524    & AGN & AGN (0.4) & AGN & AGN \\
IRAS 08572+3915      & AGN & --- & AGN & --- \\
IRAS 23060+0505      & AGN & AGN ($<$0.1) & --- & AGN \\ 
IRAS 20460+1925      & AGN & --- & --- & AGN \\
\hline
\end{tabular}
\end{center}

Note. --- 
Col. (1): Object name.
Col. (2): Energy source derived by this work. 
          ``AGN'' and ``SB'' mean AGN-dominated and 
          starburst-dominated, respectively.
          ``SB+AGN'' means both kinds of activity contribute 
          roughly equally.
Col. (3): Energy source derived based on the ratio between 
          the 7.7 $\mu$m PAH peak flux and 7.7 $\mu$m continuum flux.
          The ratios are shown in parentheses. 
          Adopted from Rigopoulou et al. (1999) except for 
          IRAS 05189$-$2524 for which the ratio is adopted from 
          Laureijs et al. (2000).
          The average ratios for starburst galaxies and AGNs are 
          $\sim$3.0 and $\sim$0.2, respectively 
          (Rigopoulou et al. 1999). 
          Those with the ratios of $>$1.0 and $<$1.0 are classified  
          as ``AGN'' and ``SB'', respectively 
          (Rigopoulou et al. 1999). 
Col. (4): Energy source derived from high spatial resolution 
          8--25 $\mu$m imaging (Soifer et al. 1999).
Col. (5): Energy source derived from 2 $\mu$m spectroscopy 
          (Veilleux et al. 1999b).
          Those with a detected broad emission line 
          whose luminosity is high enough to account for the 
          far-infrared luminosity by AGN activity are classified 
          as ``AGN''.      

\normalsize

\clearpage

\figcaption[]{ 
Expected 3--4 $\mu$m spectral shapes.
($a$) Sources powered by starburst activity. 
     PAH emission is expected to be detected at 3.3 $\mu$m 
     and sometimes at 3.4 $\mu$m.
     The effect of dust extinction reduces the PAH and continuum fluxes, 
     but does not change the equivalent width of the 3.3 $\mu$m PAH 
     emission feature significantly because flux attenuation of the 
     3.3 $\mu$m PAH and 3--4 $\mu$m continuum emission is similar.
($b$) Sources powered by obscured AGN activity. 
     The 3.4 $\mu$m carbonaceous dust absorption is expected to be 
     detected.
($c$) Sources powered by unobscured AGN activity.
     The 3--4 $\mu$m spectrum should be featureless.
($d$) Sources powered both by starburst and unobscured AGN activity.
     The 3.3 $\mu$m PAH emission feature is expected to be detected, 
     but its equivalent width should be smaller than those of 
     starburst-dominated galaxies 
     because of the contribution of AGN activity to the 3--4 $\mu$m 
     continuum emission. 
($e$) Sources powered by both starburst and obscured AGN activity.
     The 3.3 $\mu$m PAH emission feature is expected to be detected.
     When a significant fraction of the 3--4 $\mu$m continuum emission 
     originates in the obscured AGN activity, then 
     the equivalent width of the 3.3 $\mu$m PAH emission feature 
     is smaller than those of starburst-dominated galaxies, and 
     the 3.4 $\mu$m carbonaceous absorption may be detectable 
     if this feature is not veiled by possible PAH emission 
     at 3.4 $\mu$m.
     However, when emission from the obscured AGN activity is so highly 
     attenuated that its contribution to the 3--4 $\mu$m continuum 
     emission is very small, the apparent spectral shape is almost 
     the same as those powered only by starburst activity.
\label{fig1}} 

\figcaption[]{ 
Flux-calibrated spectra.
The abscissa is the observed wavelength in $\mu$m and 
the ordinate is the flux in $F_{\lambda}$.
The two down arrows indicate, respectively, the expected wavelengths of 
the redshifted 3.3 $\mu$m PAH emission (rest-frame 3.29 $\mu$m) 
and 3.4 $\mu$m absorption (rest-frame 3.4 $\mu$m) features. 
The solid lines in the spectra of Mrk 231, IRAS 05189$-$2524, 
IRAS 08572+3915, IRAS 23060+0505, and IRAS 20460+1925 are 
adopted continuum levels to investigate the 3.4 $\mu$m 
carbonaceous dust absorption feature.
Pf$\gamma$ emission lines are shown in the spectra of IC 694 and NGC 253.
In the spectrum of NGC 6240, the H$_{2}$ 0--0 S(15) emission line 
is shown.
To reduce the scatter of data points in the continuum, 
spectra are shown with the spectral resolution of $\sim$380 
except for IRAS 23060+0505 and IRAS 20460+1925 where it is $\sim$200.
\label{fig2}} 


\clearpage

\begin{thebibliography}{} 
\bibitem{}Alonso-Herrero, A., Rieke, G. H., Rieke, M. J., \& 
          Scoville, N. Z. 2000, ApJ, 532, 845
\bibitem{}Armus, L., Heckman, T. M., \& Miley, G. K. 1989, ApJ, 347, 727
\bibitem{}Blain, A. W., Kneib, J. -P., Ivison, R. J., \& Smail, I.  
          1999, ApJ, 512, L87
\bibitem{}Brand, P. W. J. L., Moorhouse, A., Burton, M. G., 
          Geballe, T. R., Bird, M., \& Wade, R. 1988, ApJ, 334, L103
\bibitem{}Bridger, A., Wright, C. S., \& Geballe, T. R. 1994, 
          in McLean I., ed., 
          Infrared Astronomy with Arrays: The Next Generation, 
          Kluwer, Dordrecht, p.537
\bibitem{}Clavel, J., et al. 2000, A\&A, 357, 839 
\bibitem{}Dennefeld, M., \& Desert, F. X. 1990, A\&A, 227, 379
\bibitem{}Dudley, C. C. 1999, MNRAS, 307, 553
\bibitem{}Dudley, C. C., Wynn-Williams C. G. 1997, ApJ, 488, 720 
\bibitem{}Elbaz, D., et al. 1998, Proceedings 34th Liege 
          International Astrophysics Colloquium 
          `The Next Generation Space Telescope: Science Drivers and 
          Technological Challenges' Liege, Belgium, p.47, 
          (astro-ph/9807209)
\bibitem{}Engelbracht, C. W., Rieke, M. J., Rieke, G. H., Kelly, D. M., 
          \& Achtermann, J. M. 1998, ApJ, 505, 639 
\bibitem{}Frogel, J. A., Gillett, F. C., Terndrup, D. M., \& 
          Vader, J. P. 1989, ApJ, 343, 672
\bibitem{}Genzel, R., \& Cesarsky, C. J. 2000, ARA\&A, in press  
          (astro-ph/0002184)  
\bibitem{}Genzel, R., Weitzel, L., Tacconi-Garman, L. E., Blietz, M., 
          Cameron, M., Krabbe, A., \& Lutz, D. 1995, ApJ, 444, 129
\bibitem{}Genzel, R., et al. 1998, ApJ, 498, 579
\bibitem{}Glass, I. S., \& Moorwood, A. F. M. 1985, MNRAS, 214, 429
\bibitem{}Goldader, J. D., Joseph, R. D., Doyon, R., \& Sanders, D. B.  
          1995, ApJ, 444, 97 
\bibitem{}Heisler, C. A., Norris, R. P., Jauncey. D. L., Reynolds, J. E., 
          \& King, E. A. 1998, MNRAS, 300, 1111 
\bibitem{}Hill, G. J., Wynn-Williams, C. G., \& Becklin, E. E. 1987, 
          ApJ, 316, L11
\bibitem{}Imanishi, M., \& Ueno, S. 2000, ApJ, 535, 626 
\bibitem{}Imanishi, M., Terada, H., Goto, M., \& Maihara, T. 1998, 
          PASJ, 50, 399
\bibitem{}Imanishi, M., Terada, H., Sugiyama, K., Motohara, K., 
          Goto, M., \& Maihara, T. 1997, PASJ, 49, 69  
\bibitem{}Ivezic, Z., Nenkova, M., \& Elitzur, M. 1999, 
          User Manual for DUSTY, University of Kentucky Internal Report, 
          accessible at http://www.pa.uky.edu/$^{\sim}$moche/dusty/  
          (astro-ph/9910475)
\bibitem{}Joy, M., Telesco, C. M., Decher, R., Lester, D. F., 
          Harvey, P. M., Rickard, L. J., \& Bushouse, H. 1989, ApJ, 
          339, 100
\bibitem{}Kalas, P., \& Wynn-Williams, C. G. 1994, ApJ, 434, 546
\bibitem{}Kewley, L. J., Heisler, C. A., Dopita, M. A., 
          Sutherland, R., Norris, R. P., Reynolds, J., \& 
          Lumsden, S. 2000, ApJ, 530, 704
\bibitem{}Kim, D. -C., \& Sanders, D. B. 1998, ApJS, 119, 41 
\bibitem{}Laureijs, R. J., et al. 2000, A\&A, in press  
          (astro-ph/0005076)
\bibitem{}Lonsdale, C. J., Lonsdale, C. J., \& Smith, H. E.  
          1992, ApJ, 391, 629
\bibitem{}Lonsdale, C. J., Smith, H. E., \& Lonsdale, C. J. 1993, 
          ApJ, 405, L9
\bibitem{}Lutz, D., et al. 1996, A\&A, 315, L269
\bibitem{}Lutz, D., Spoon, H. W. W., Rigopoulou, D., Moorwood, A. F. M., 
          Genzel, R. 1998, ApJ, 505, L103 
\bibitem{}Maloney, P. R. 1999, Ap\&SS, in press, 
          The Ringberg Workshop, 
          ``Ultraluminous Galaxies: Monsters or Babies''   
          (astro-ph/9903275) 
\bibitem{}Mathis, J. S. 1990, ARA\&A, 28, 37
\bibitem{}Moorwood, A. F. M. 1986, A\&A, 166, 4
\bibitem{}Mountain, C. M., Robertson, D. J., Lee, T. J., \& Wade, R.  
          1990, Proc. SPIE, 1235, 25
\bibitem{}Mouri, H., Kawara, K., Taniguchi, Y., \& Nishida, M. 
          1990, ApJ, 356, L39
\bibitem{}Pendleton, Y. 1996, in The Cosmic Dust Connection, 
          Greenberg, J. M., ed., Kluwer Academic Publishers, p.71
\bibitem{}Pendleton, Y. J., Sandford, S. A., Allamandola, L. J., 
          Tielens, A. G. G. M., \& Sellgren, K. 1994, ApJ, 437, 683 
\bibitem{}Rice, W., Lonsdale, C. J., Soifer, B. T., Neugebauer, G., 
          Kopan, E. L., Lloyd, L. A., De Jong, T., \& Habing, H. J.  
          1988, ApJS, 68, 91
\bibitem{}Rieke, G. H., \& Lebofsky, M. J. 1985, ApJ, 288, 618
\bibitem{}Rieke, G. H., Curti, R. M., Black, J. H., Kailey, W. F., 
          McAlary, C. W., Lebofsky, M. J., \& Elston, R. 1985, ApJ, 
          290, 116
\bibitem{}Rieke, G. H., Lebofsky, M. J., Thompson, R. I., Low, F. J., 
          \& Tokunaga, A. T. 1980, ApJ, 238, 24
\bibitem{}Rigopoulou, D., Spoon, H. W. W., Genzel, R., Lutz, D., 
          Moorwood, A. F. M., \& Tran, Q. D. 1999, AJ, 118, 2625
\bibitem{}Roche, P. F., Aitken, D. K., Smith, C. H., \& Ward, M. J.  
          1991, MNRAS, 248, 606 
\bibitem{}Rudy, R. J., Levan, P. D., \& Rodriguez-Espinosa, J. M.  
          1982, AJ, 87, 598
\bibitem{}Sanders, D. B., \& Mirabel, I. F. 1996, ARA\&A, 34, 749
\bibitem{}Satyapal, S., Watson, D. M., Forrest, W. J., Pipher, J. L., 
          Fischer, J., Greenhouse, M. A., Smith, H. A., \& 
          Woodward, C. E. 1999, ApJ, 516, 704
\bibitem{}Schmitt, H. R., Kinney, A. L., Calzetti, D., \& 
          Bergmann, T. S. 1997, AJ, 114, 592 
\bibitem{}Scoville, N. Z., et al. 2000, AJ, 119, 991
\bibitem{}Smith, H. E., Lonsdale, C. J., \& Lonsdale, C. J. 1998, 
          ApJ, 492, 137
\bibitem{}Soifer, B. T., Sanders, D. B., Madore, B. F., Neugebauer, G., 
          Danielson, G. E., Elias, J. H., Lonsdale, C. J., \& 
          Rice, W. L. 1987, ApJ, 320, 238 
\bibitem{}Soifer, B. T., et al. 2000, AJ, 119, 509
\bibitem{}Sugai, H., Malkan, M. A., Ward, M. J., Davies, R. I., \&  
          McLean, I. S. 1997, ApJ, 481, 186 
\bibitem{}Sugai, H., Davies, R. I., Malkan, M. A., McLean, I. S., 
          Usuda, T., \& Ward, M. J. 1999, ApJ, 527, 778
\bibitem{}Surace, J. A., \& Sanders, D. B. 1999, ApJ, 512, 162
\bibitem{}Surace, J. A., Sanders, D. B., \& Evans, A. S. 2000, ApJ, 
          529, 170
\bibitem{}Tokunaga, A. T., Sellgren, K., Smith, R. G., Nagata, T., 
          Sakata, A., \& Nakada, Y. 1991, ApJ, 380, 452 
\bibitem{}Tokunaga, A. T. 2000, in Allen's Astrophysical Quantities, ed. 
          A. N. Cox (4th ed: AIP Press: Springer), Chapter 7, p.143 
\bibitem{}Tully, R. B. 1988, Nearby Galaxies Catalog, 
          Cambridge University Press, Cambridge, p.10 
\bibitem{}Vader, J. P., Frogel, J. A., Terndrup, D. M., \& 
          Heisler, C. A. 1993, AJ, 106, 1743 
\bibitem{}Veilleux, S., Kim, D. -C., \& Sanders, D. B. 1999a, 
          ApJ, 522, 113
\bibitem{}Veilleux, S., Kim, D. -C., Sanders, D. B., Mazzarella, J. M., 
          \& Soifer, B. T. 1995, ApJS, 98, 171 
\bibitem{}Veilleux, S., Sanders, D. B., \& Kim, D. -C. 1999b, ApJ, 
          522, 139
\bibitem{}Vignati, P., et al. 1999, A\&A, 349, L57
\bibitem{}Voit, G. M. 1992, MNRAS, 258, 841
\bibitem{}Wright, G. S., Bridger, A., Geballe, T. R., \& 
          Pendleton, Y. 1996, 
          New Extragalactic Perspectives in the New South Africa, 
          Kluwer Academic Publishers, Block, D. L., \& Greenberg, J. M., 
          (eds), p.143
\bibitem{}Zhou, S., Wynn-Williams, C. G., \& Sanders, D. B. 1993, 
          ApJ, 409, 149
\end{thebibliography}
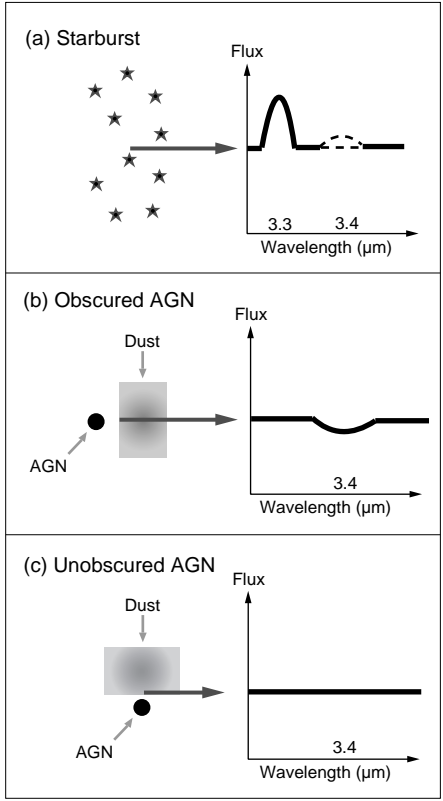
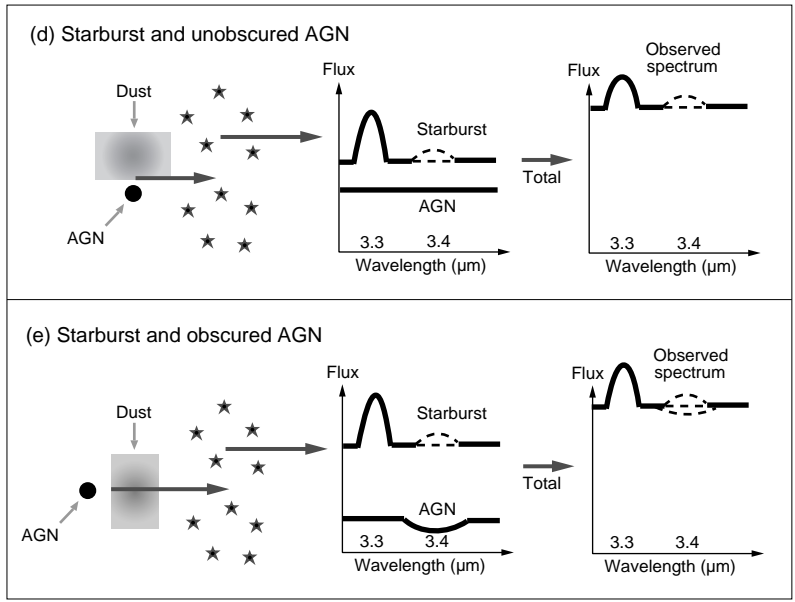
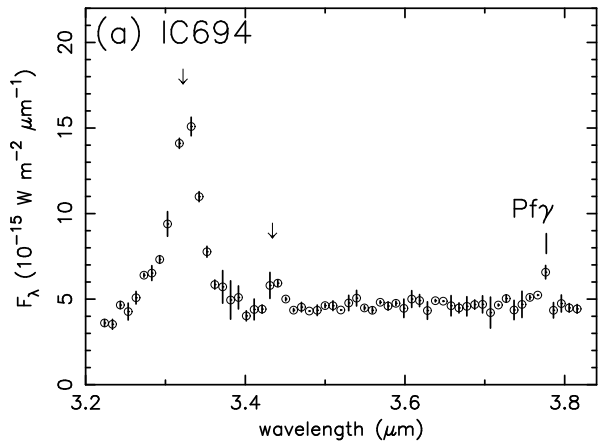
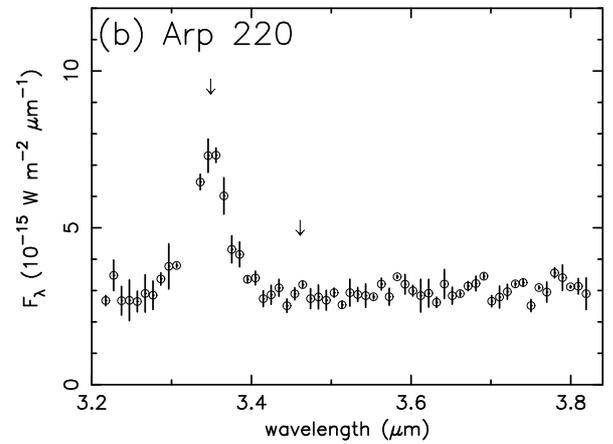
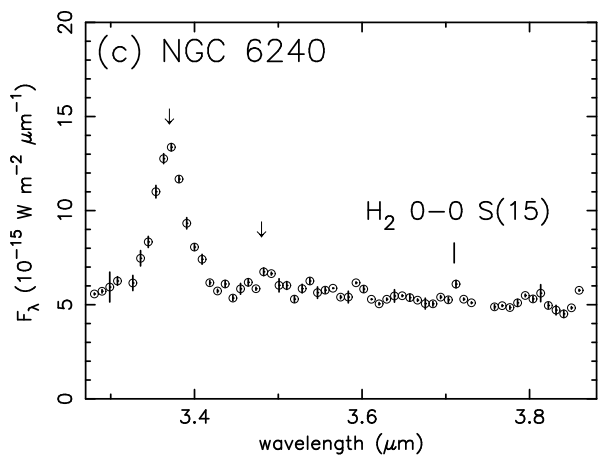
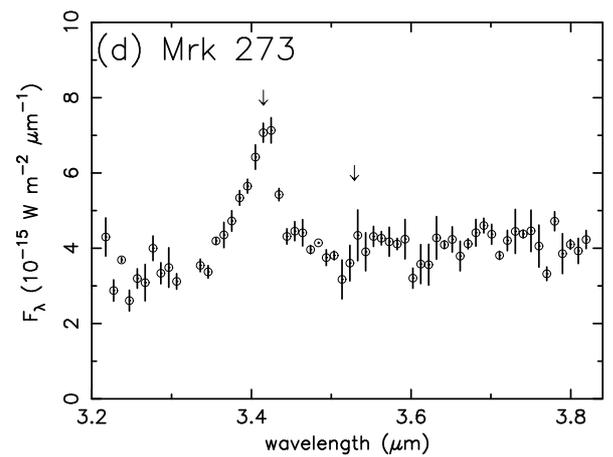
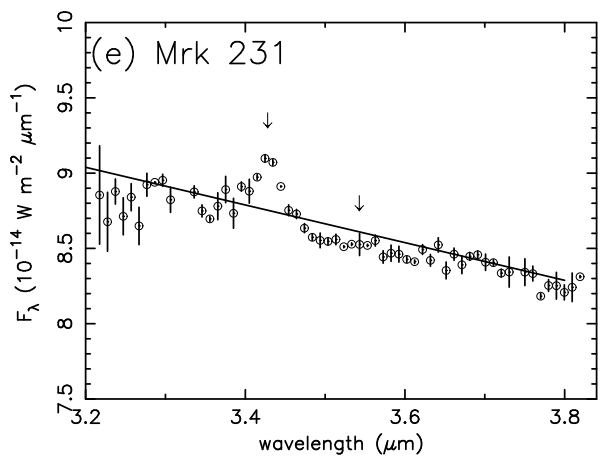
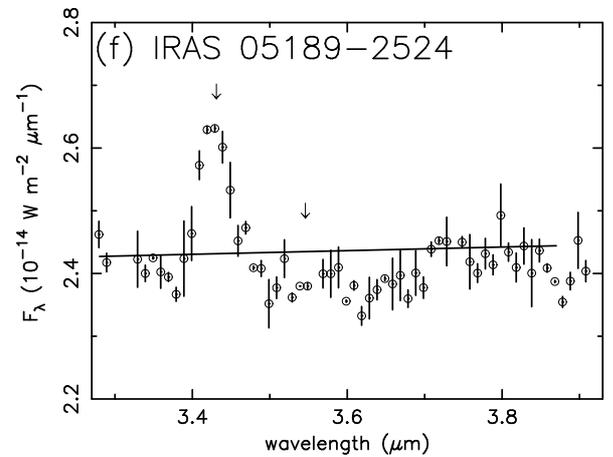
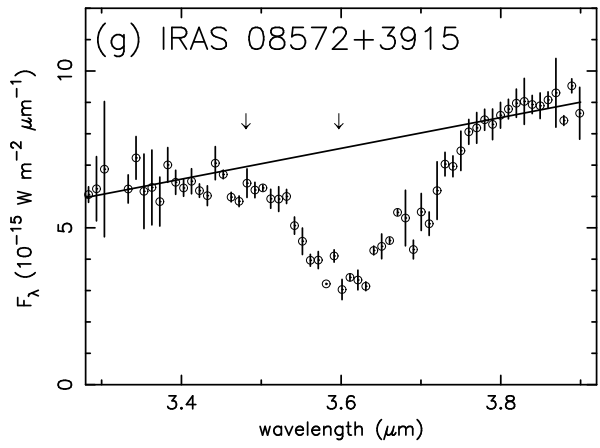
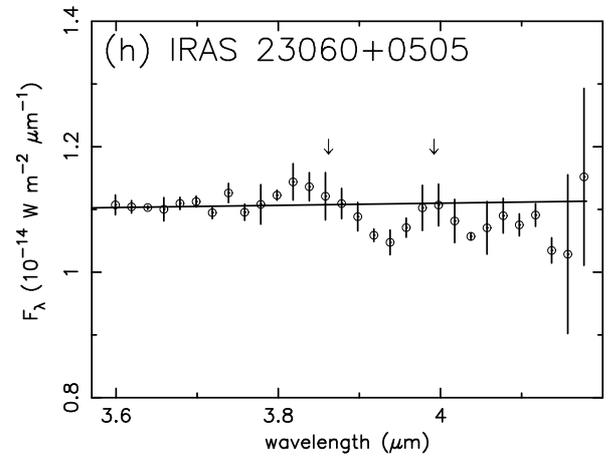
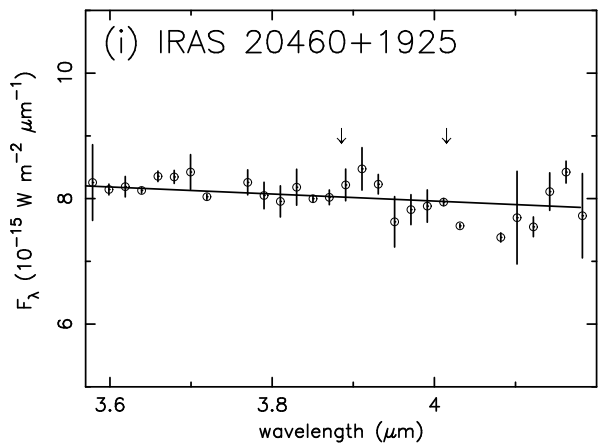
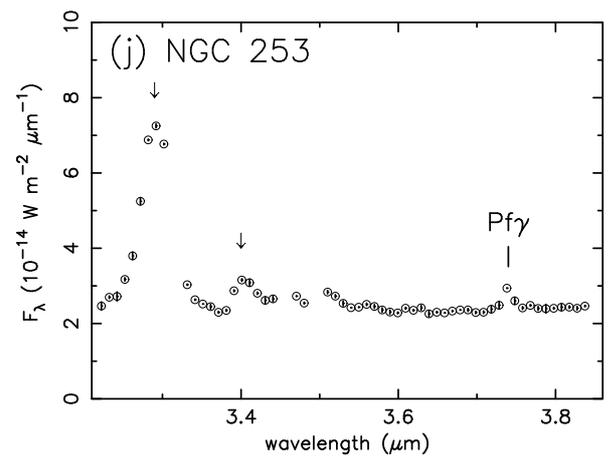
\end{document}